\documentstyle[12pt,epsf]{article}
\advance\textheight by \topskip
\begin{document}
\thispagestyle{empty}
\begin{flushright}
KUCP0143\\
Nov. 1, 1999\\
\end{flushright}
\vskip 2 cm
\begin{center}
{\LARGE\bf An Open Universe from Valley Bounce}
\vskip 1.7cm

 {\bf Kazuya Koyama}\footnote{
E-mail: kazuya@phys.h.kyoto-u.ac.jp} 
{\bf Kayoko Maeda}
\footnote{E-mail: maeda@phys.h.kyoto-u.ac.jp }
{\bf Jiro Soda}
\footnote{E-mail: jiro@phys.h.kyoto-u.ac.jp } \

\vskip 1.5mm
   
\vskip 2cm
 $^1$ Graduate School of Human and Environment Studies, Kyoto University, 
       Kyoto  606-8501, Japan \\
  $^2$ $^3$ Department of Fundamental Sciences, FIHS, Kyoto University,
       Kyoto, 606-8501, Japan \\
\end{center}

{\centerline{\large\bf Abstract}}
\begin{quotation}
\vskip -0.4cm
It appears difficult to construct a simple model
for an open universe based on the one bubble inflationary scenario.
The reason is that one needs a large mass to avoid the
tunneling via the Hawking Moss solution and a small mass for successful 
slow-rolling. However, Rubakov and  Sibiryakov suggest that the 
Hawking Moss solution is not a solution for the false vacuum decay 
process because it does not satisfy the boundary condition. 
Hence, we have reconsidered the arguments for the defect of 
the simple polynomial model. 
We point out the possibility that one of the valley bounce belonging
to a valley line in the functional space represents the decay process
instead of the Hawking Moss solution. 
Under this presumption, we show an open inflation model 
can be constructed within the polynomial form of the potential so
that the fluctuations can be reconciled with the observations. \\
\hspace{1cm}\\
PACS: 98.80.Cq $\quad \quad$ Keywords: One-bubble open universe, 
Valley method

\end{quotation}
 \newpage

\section{Introduction}
\hspace{0.1cm}

Recent observations suggest the matter density of the universe is 
less than the critical density. 
Hence, it is desirable to have a model for an open universe, 
say $\Omega_0 \sim 0.3$. The realization of an open universe is
difficult in the ordinary inflationary scenario. This is because
if the universe expands enough to solve the horizon problem, 
the universe becomes almost flat. One attempt to realize 
an open universe in the inflationary scenario is to consider inside 
the bubble created by the false vacuum decay \cite{Gott}. 
The scenario is as follows. Consider the
potential which has two minimum. One is the false vacuum which has 
non-zero energy and the other is the true vacuum. Initially the field is 
trapped at the false vacuum. Due to the potential energy, universe expands
exponentially and the large fraction of the universe becomes
homogeneous. As the false vacuum is unstable, it decays and creates 
the bubble of the true vacuum. If the decay process is well suppressed, 
the interior of the bubble is still homogeneous. 
The decay is described by the $O(4)$ symmetric configuration 
in the Euclidean spacetime.
Then, analytical continuation of this configuration 
to the Lorentzian spacetime 
describes the evolution of the bubble which looks from the inside like
an open universe.
Unfortunately, since the bubble radius cannot be greater than the
Hubble radius, the created universe is curvature dominated even if the
whole energy of the false vacuum is converted to the energy of the matter
inside the bubble \cite{Yokoyama1}. 
Thus, the second inflation in the bubble is needed. 
If this second inflation stopped when $\Omega<1$, our universe becomes 
homogeneous open universe. 

Though the basic idea is simple, the realization of the scenario
in a simple model has been recognized difficult \cite{Linde1}.
The difficulty is usually explained as follows.
Consider the model involving one scalar field. 
For the polynomial form of the potential like 
$V(\phi)= m^2\phi^2- \delta \phi^3 + \lambda \phi^4$, 
the tunneling should occur at sufficiently large $\phi$ 
to ensure that the second inflation gives the appropriate density parameter. 
Then, the curvature around the barrier which separates the false and
the true vacuum is small compared with the Hubble
scale which is determined by the energy of the false vacuum. 
The field jumps up onto the top of the barrier due to the quantum diffusion.  
When the field begins to roll down from the top of the 
barrier, large fluctuations are formed due to the quantum diffusion at
the top of the barrier. Then the whole scenario fails. The problem is
rather generic. To avoid jumping up, the curvature around the barrier
should be large compared with the Hubble scale $V''> H^2$. On the other
hand, to realize the second inflation, the field should roll down 
slowly, then we need $V''<H^2$. These two conditions are incompatible.

There are several attempts to overcome the problem. 
Recently Linde constructs the potential which has sharp peak near the 
false vacuum \cite{Linde2}. In this potential, 
the tunneling occurs and at the same time slow-rolling is allowed after 
the tunneling, then the second inflation can be realized. 
But, it is still unclear what is the physical mechanism for the appearance 
of the sharp peak in the potential. 

A more detailed study of the tunneling process is needed 
to tackle the problem. In the imaginary-time path-integral formalism, the 
tunneling is described by the solution of the Euclidean field equation. 
The solution gives the saddle-point of the path-integral. 
Then the solution determines the semi-classical exponent of the
decay rate $\exp(-S_E(\phi_B))$, where $S_E$ is the Euclidean action. 
In the case the curvature around the barrier is small compared with the
Hubble, the solution is given by the Hawking Moss (HM) solution, 
which stays at the top of the barrier through the whole Euclidean time 
\cite{HM}. 
Recently Rubakov and  Sibiryakov give the interpretation
of the HM solution in the de Sitter spacetime 
using the constrained instanton method \cite{Rubakov,Affleck}. 
They show the HM solution does not represent
the false vacuum decay. This is because the HM solution does not satisfy the
boundary condition that the field exists in the false vacuum at the
infinite past. One should consider a family of the almost saddle-point 
configurations instead of the true solution of the Euclidean
field equation.  They show although the decay rate is determined by the HM
solution, the structure of the field after tunneling is determined by
the other configuration which is one of the almost saddle-point solutions.
In the constrained instanton method, one introduces the constraint to 
define the subspace in the functional space and looks for the almost 
saddle-point solution. One must choose the constraint so that the almost 
saddle-point solution is contained in the region which is expected to 
dominate the path-integral. 
One way is to consider the valley region of the functional space 
\cite{Balitsky, Aoyama1}.  
Along the valley line, the action varies most gently. 
Then it is reasonable to take a
configuration on the valley line as the almost saddle-point
configuration. We will call the configuration on the valley line the 
valley bounce $\phi_V$.

Their analysis opens up a new possibility to overcome the problem. 
Suppose that one of the valley bounces 
describes the tunneling and the inside of the bubble created from this
valley bounce is our open universe. 
The initial condition of the tunneling field in the open universe 
is determined by the valley bounce.
If the field appears sufficiently far from the top at the nucleation, 
the large fluctuations can be avoided. During the tunneling,
fluctuations of the tunneling
field are generated. These fluctuations are stretched during the second
inflation and observed in the open universe.  

In this paper, 
we extend the valley method developed by Aoyama.et.al \cite{Aoyama1} 
to the de Sitter spacetime. 
Taking the assumption that 
one of the valley bounces describes the tunneling and
inside the bubble created from the valley bounce is our universe,
the fluctuations generated during the tunneling are calculated 
by defining the fluctuations which are relevant to the observable in the
open universe as those orthogonal to the gradient of the action.
Then we show an open universe can be constructed from the valley bounce 
and the fluctuations can be generated with the appropriate properties. 
The paper is organized as follows. In the next section we
review the formalism to describe the false vacuum decay
in the de Sitter spacetime. Then we explain the role of the
configurations on the valley of the action, i.e. valley bounces.
We derive the valley equation which determines the valley bounce. 
In section 3, we solve the valley equation analytically using the 
piece-wise quadratic potential and fixed background approximation.
Then the structure of the valley is shown. In section 4,
we explain the scenario emphasizing the role of the valley bounce
and show that a simple model for an open universe 
can be constructed  
without introducing the fine-tuning of the potential. 
In section 5, we calculate observable fluctuations in the
open universe from the valley bounce.
The power spectrum of the curvature perturbations 
is calculated and these fluctuations 
are shown to be compatible with the observations. 
In section 6, we summarize the results.
 
\section{Valley method in de Sitter spacetime}
\hspace{1cm}

First we review the formalisms which are necessary to describe the 
false vacuum decay in the de Sitter space.
We want to examine the case 
in which the gravity comes to play a role. Unfortunately, 
we have not known how to deal with quantum gravity effect yet. 
So, we study the case in which we can treat gravity at the semi-classical 
level. That is, we treat the problem within the framework of the field 
theory in a fixed curved spacetime \cite{Rubakov}. 
The potential relevant to the tunneling is given by
\begin{equation}
V(\phi)=\epsilon+V_T(\phi).
\end{equation}
We assume $\epsilon$ is of the order 
$M_{\ast}^4$ and $V_T(\phi)$ is of the order $M^4$. 
We study the case $M$ is small compared to $M_{\ast}$,  $M \ll  M_{\ast}$.
Then the geometry of the spacetime is fixed to the de Sitter spacetime
with $H=M_{\ast}^2/M_p$ ,where $M_p^{-2}=8 \pi G/3$.
We consider the situation in which the potential $V_T(\phi)$ has the false 
vacuum at $\phi=\varphi_F$ and the top of the barrier at $\phi=\varphi_T$. 
Since the background metric is fixed, we can change 
the origin of the energy freely. We choose $V_T(\phi_F)=0$. 
Following, we work in units with $H=1$.

The decay rate is given by the imaginary part of the path-integral
\begin{equation}
Z=\int [d \phi] \exp {\left(-S_E(\phi) \right)},
\end{equation}
where $S_E$ is the Euclidean action relevant to the tunneling.
The dominant contribution of this path-integral is given by 
the configurations which have $O(4)$ symmetry \cite{CL}. So, we assume the 
background metric and the field to have the form
\begin{eqnarray}
ds^2 &=& d \sigma^2+a(\sigma)^2 \left( d\rho^2+\sin^2 \rho
d \Omega \right), \nonumber\\
\phi &=& \phi(\sigma),
\end{eqnarray}
where $a(\sigma)=\sin \sigma$. 
Then, the Euclidean action of $\phi(\sigma)$ is given by
\begin{equation}
S_E = 2 \pi^2 \int d \sigma    \left( a^3
\left( \frac{1}{2}\phi'^2+V_T(\phi) \right) \right).
\end{equation}  

The saddle-point of this path-integral is determined by 
the Euclidean field equation $\delta S_E/\delta \phi$=0;
\begin{equation}
\phi''+3 \cot \sigma \: \phi'-V_T'(\phi)=0.
\end{equation}
We impose the regularity conditions at the time when $a(\sigma)=0$ as
\begin{equation}
\phi'(\sigma=0)=\phi'(\sigma=\pi)=0.
\end{equation}
We represent the solution of this equation as $\phi_B(\sigma)$.
If the fluctuations around the solution have a negative mode, it
gives the imaginary part to the path-integral and this solution
contributes to the decay dominantly. The decay rate $\Gamma$ is evaluated by
\begin{equation}
\Gamma \sim \exp (-S_E(\phi_B)).
\end{equation}
The equation has two types of the solutions depending on the shape of 
the potential.
If the curvature around the barrier is large compared with the Hubble
scale, then the Coleman De Luccia (CD) solution  
and the Hawking Moss (HM) solution exist \cite{HM,CL}. 
Since the CD solution has lower action than that of the HM solution, 
the decay is described by the CD solution. The analytic
continuation of this solution to Lorentzian spacetime describes the
bubble of the true vacuum. On the other hand, in the case the curvature 
around the barrier is small compared with the Hubble scale, 
only the HM solution exists. This solution is a trivial solution 
$\phi=\varphi_T$. The meaning of the HM solution is somewhat ambiguous. 
There are several attempts to interpret this tunneling mode. 
One way is to use the stochastic approach \cite{Linde3}. 
It has been demonstrated that the decay rate given by eq.(7) coincides 
with the probability of jumping from the false vacuum $\varphi_F$ onto the
top of the barrier $\varphi_T$ due to the quantum fluctuations. 

Recently, Rubakov and Sibiryakov gave the interpretation of the 
HM solution using the constrained instanton method
\cite{Rubakov,Affleck}. The main idea is to
consider a family of the almost saddle-point configurations instead
of the true solution of the Euclidean field equation, i.e. the HM solution. 
The motivation comes from the boundary condition. They take the boundary 
condition that the state of the quantum fluctuations above the classical 
false vacuum is the conformal vacuum. For this boundary condition 
they show the field should not be constant at $0 < \sigma < \pi$ and the HM
solution is excluded by this boundary condition. 
Then one should seek the other configurations which obey the boundary
condition and dominantly contribute to the path-integral.
In the functional integral, the saddle-point solution gives the most
dominant contribution, but the contribution from a family of almost
saddle-point configurations which have almost the same action with
that of the saddle-point solution should also be included. 
To seek the almost saddle-point solution, one introduces the 
constraint to the path integral. The constraint selects the subspace 
of the functional space. 
The minimum in this subspace satisfies the equation of motion 
with constraint instead of the field equation. This minimum corresponds
to the almost saddle-point configuration
which is slightly deformed from the HM solution.  
Since the HM solution gives the minimum action, 
the decay rate is determined by the HM solution. But the
structure of the field after tunneling can be determined 
by the one of the almost saddle-point configuration.
They found that the configuration describes the bubble of the true vacuum 
in the Lorentzian spacetime. 
Then, they conclude that even 
in the case only the HM solution exists, the result of the tunneling 
process can be the bubble of the true vacuum which is described by one of
the almost saddle-point configurations.

In the constrained instanton method, the validity of the method depends on 
the choice of
the constraint \cite{Aoyama1,Aoyama2}. 
We should choose the constraint so that the almost saddle-point solution
is contained in the region 
which is expected to dominate the path-integral. 
Since the action varies most gently along
the valley line, we consider the valley region of the action 
\cite{Balitsky, Aoyama1}.  

In the false vacuum decay process in the flat spacetime, the configurations
along the valley line have physical meanings \cite{Aoyama3}. These
configurations actually dominate the path-integral in the presence of the
low energy incoming particles. If the incoming particles exist, the
path-integral which describes the decay is given by $\int [d \phi]
\phi^q \exp(-S_E(\phi))$. The effect of the incoming particles deforms
the saddle-point. As long as the energy of these incoming particles
is low, the deformed configurations belong to the valley. This
is because in deforming the configurations from the saddle-point solution, 
the configurations on the valley line can be obtained most easily compared 
with the other configurations with the same action. 
Thus, the configurations on the valley play a crucial role to
calculate the decay rate or the cross section when 
the initial state is of higher energy than the ground state.
In the de Sitter space time,
the choice of the quantum states of the quantum fluctuations
above a given classical false vacuum
will affect the tunneling process considerably \cite{Rubakov}. 
Hence, we think the
configurations on the valley line in the de Sitter spacetime may also play an 
important role, though it is not easy to identify the corresponding 
quantum state. 

Taking into account the above fact, it is desirable to analyze the 
structure of not only the solution of the Euclidean field equation
but also the configurations on the valley line.
One way to fined the configurations on this valley line is
to use the valley method developed by Aoyama.et.al \cite{Aoyama1}.
To obtain the intuitive understanding of this method, consider the
system of the field $\phi_i$. Here $i$ stands for the discretized
coordinate label and we take the metric as $\delta_{ij}$. 
In the valley method the equation which identifies the valley
line in the functional space is given by
\begin{equation}
D_{ij} \partial_i S = \lambda \: \partial_i S,\:\:\:\: D_{ij}=\partial_i 
\partial_j S,
\end{equation}
where $\partial_i=\partial/\partial \phi_i$. Since the equation (8) has 
one parameter $\lambda$, the equation defines a trajectory in the space of
$\phi$. The parameter $\lambda$ is one of the eigen value of the matrix
$D_{ij}$. On the trajectory the gradient vector $\partial_i S$ is
orthogonal to all the eigenvectors of $D_{ij}$ except for the
eigenvector of the eigen value $\lambda$.
The equation can be rewritten as
\begin{equation}
\partial_i \left( \frac{1}{2}(\partial_j S)^2-\lambda S \right)=0.
\end{equation}
Then the solution extremizes the norm of the gradient vector $\partial_i S$ 
under the constraint $S=$const., where
$\lambda$ plays the role of the Lagrange multiplier. Such solution
can be found each hypersurface of constant action, then the solutions
of the equation form a line in the functional space. 
If we take $\lambda$ as the 
one with the smallest value, then the gradient vector is minimized
and the action varies most gently along this line. 
This is a plausible definition of the valley line. 
We will call the configuration on the valley line of the
action the valley bounce $\phi_V$ and the trajectory they form the valley
trajectory. 

We shall formulate the valley method in the de Sitter spacetime.
The most convenient way is to use the variational method eq.(9).
We shall define the valley action by
\begin{equation}
S_V = S_E-\frac{1}{2 \lambda} \int d \sigma  \sqrt{g} 
\left( \frac{1}{\sqrt{g}} \frac{\delta S_E}{\delta \phi} \right)^2.
\end{equation}
The valley bounce is obtained by varying the action $S_V$.
The equation which determines the valley bounce $\delta S_V/\delta \phi=0$
is a fourth order differential equation. 
We introduce the auxiliary field $f$ to cancel the fourth derivative
term \cite{Aoyama3};
\begin{equation}
S_{f}=\frac{1}{2 \lambda} \int d \sigma \sqrt{g}
\left( f-\frac{1}{ \sqrt{g}} \frac{\delta S_E}{\delta \phi} \right)^2.
\end{equation}
Then the valley action becomes
\begin{equation}
S_V+S_f = S_E+\frac{1}{2 \lambda} \int d \sigma \sqrt{g} f^2 
-\frac{1}{\lambda} 
\int d \sigma f \frac{\delta S_{E}}{\delta \phi}.
\end{equation}
Taking the variation of this action with respect to $f$ and $\phi$, we obtain
the equations for $\phi$ and $f$;
\begin{eqnarray}
\frac{1}{\sqrt{g}} \frac{\delta S_E}{\delta \phi} &=& f,
\nonumber\\
\int d \sigma'
\frac{\delta^2 S_E}{\delta \phi(\sigma) \delta \phi(\sigma')} f(\sigma)
&=& \lambda  \sqrt{g} f(\sigma).
\end{eqnarray}
Using $a(\sigma)=\sin \sigma$, the valley equation 
which determines the structure of the valley bounce is given by
\begin{eqnarray}
\phi''+3 \cot  \sigma \: \phi'- V_T'(\phi) &=& -f,  \nonumber\\
f''+3 \cot  \sigma  \: f'-V_T''(\phi)f &=& - \lambda f.
\end{eqnarray}
The fluctuations around the valley bounce can be expanded by 
the eigenmodes $g_{\alpha, n}(\sigma)$ 
with the eigenvalue$  \rho_{\alpha, n}$ of the operator 
$(\delta S_E^2/\delta \phi \delta \phi)_{\phi_{\alpha}}$;
\begin{equation}
g_{\alpha, n}''+3 \cot  \sigma  \: g_{\alpha, n}'-V_T''(\phi)
g_{\alpha, n} = - \rho_{\alpha, n} g_{\alpha, n}.
\end{equation}
Since $\lambda$ is one of the $\rho_{\alpha, n}$ with the smallest
value,  the gradient of the action $f(\sigma)$ is orthogonal to 
the other eigenmodes with $\rho_{\alpha, n} \ne \lambda$. 
To ensure that the valley
bounce gives the imaginary part to the path-integral, the fluctuations around 
the valley bounce should have one negative eigenvalue $\rho_{-}$. 
 
In the next section, we solve the valley equation and clarify
the structure of the valley bounce.
Solving the valley equation is the eigenvalue problem of the two 
variables, it is desirable to solve the equation analytically to
confirm the existence of the solutions. The valley bounce can have 
a thick-wall profile. In the flat spacetime, there exists the 
attempt to treat thick-wall solutions analytically 
by constructing the piece-wise quadratic potentials \cite{Hamazaki}.
In the next section, we extend the attempt to the de Sitter 
spacetime and solve not only the Euclidean field equation 
but also the valley equation analytically. 

\section{Valley bounces}
\subsection{The construction of Valley bounces}
To solve the valley equation developed in the last section, 
we construct the piece-wise quadratic potential. 
We connect two parabola. In the potential the true vacuum is absent.
But this is not essential in calculations. 
In fact we have solved numerically the valley equation for several 
potentials and found this model is sufficient to discuss the generic feature 
of the valley bounce.

The potential which we study is
\begin{eqnarray}
 V_T(\phi) = \left\{ 
\begin{array}{ll}
\frac{1}{2}m_F^2(\phi-\varphi_F)^2,&
\qquad -\infty <\phi < 0,  \\
\\
-\frac{1}{2}m_T^2(\phi-\varphi_T)^2+\eta,&
\qquad 0 \leq \phi< \infty, \end{array}
\right.
\end{eqnarray}
where $\eta$ is of the order $M^4$.
We require that the potential and its derivative are connected smoothly
at the connection point $\phi=0$. From this condition, we obtain
\begin{eqnarray}
\varphi_T &=& -\frac{m_F^2}{m_T^2} \varphi_F, \nonumber\\
\varphi_F &=& -\sqrt{\frac{2 m_T^2 \: \eta}{m_F^2(m_F^2+m_T^2)}}.
\end{eqnarray}
$\varphi_T$ and $\varphi_F$ have a mass scale of the order $M^2/m_T$.
Thus, we rescale the field as $\phi \to (M^2/m_T) \phi$ and 
$f \to (M^2/m_T) f$.

First we solve the Euclidean field equation; 
\begin{equation}
\phi''+3 \cot \sigma \: \phi'- V_T'(\phi) = 0.  
\end{equation}
If one puts $z=-\cos \sigma$ and
$Y(z)=\sqrt{1-z^2}(\phi-\varphi_i)$, the equation reduces to
the associated Legendre differential equation
\begin{equation}
(1-z^2) \frac{d^2 Y}{d z^2}-2 z \frac{d Y}{d z}+ 
\left[\nu_i (\nu_i+1)-\frac{\mu^2}{1-z^2} \right]Y=0,
\end{equation}
where $\mu=1$ and $\nu_i$ is given by
\begin{equation}
\nu_T=\sqrt{\frac{9}{4}+m_T^2}-\frac{1}{2}, \:\:\:
\nu_F=\sqrt{\frac{9}{4}-m_F^2}-\frac{1}{2}.
\end{equation}
Here, $i=T$ for $-\infty< \phi<0$ and $i=F$ for $0 \leq \phi <\infty$.
The independent solutions 
of the equation are given by the associated Legendere function of the first
and second kinds, $P^{1}_{\nu_i}(z)$ and $Q^{1}_{\nu_i}(z)$. 
$P^{1}_{\nu}(z)$ is regular at $z \to 1$. 
Since these functions behave at $z \to -1$ as
\begin{eqnarray}
P^{1}_{\nu}(z) &\to& - 2^{1/2} \sin(\pi \nu) \pi^{-1} 
(1+z)^{-1/2}, \nonumber\\
Q^{1}_{\nu}(z) &\to&  - 2^{-1/2} \cos(\pi \nu) 
(1+z)^{-1/2},
\end{eqnarray}
the combination of these solutions
\begin{equation}
B^{\mu}_{\nu}(z)=P^{\mu}_{\nu}(z)+\left(-\frac{2}{\pi} \tan(\pi \nu) \right)
Q^{\mu}_{\nu}(z)
\end{equation}
is regular at $z \to -1$.
Then the solution
which satisfies the boundary condition is given by
\begin{eqnarray}
\phi_B = \left\{ 
\begin{array}{ll}
\displaystyle{
\varphi_F+\frac{1}{\sqrt{1-z^2}} \:A_F \:B^1_{\nu_F}(z)},&
\qquad -1 \leq z <z_0, \\
\\
\displaystyle{
\varphi_T+\frac{1}{\sqrt{1-z^2}} \:A_T \:P^1_{\nu_T}(z)},&
\qquad z_0 \leq z \leq 1,\end{array}
\right.
\end{eqnarray}
where $\phi_B(z_0)=0$.
Since the potential is constructed to be smooth to its first derivative,
we demand $\phi_B$ 
and its first derivative must be continuous at $z=z_0$. Then
the coefficients $A_i$ are determined in terms of $z_0$;
\begin{equation}
A_F(z_0) = -\frac{\varphi_F \: \sqrt{1-z_0^2}}{B^1_{\nu_F}(z_0)}, \:\:\:\:\:
A_T(z_0) = -\frac{\varphi_T \: \sqrt{1-z_0^2}}{P^1_{\nu_T}(z_0)}. 
\end{equation}
The junction time $z_0$ is determined by $\phi_B(z_0)=0$;
\begin{equation}
\varphi_F P^1_{\nu_{T}}(z_0)B^2_{\nu_{F}}(z_0)- \varphi_T P^2_{\nu_{T}}
(z_0)B^1_{\nu_{F}}(z_0)=0.
\end{equation}
If the algebraic equation for $z_0$ has a solution, this gives the 
CD solution. The condition for the existence of the solution 
restricts the parameter $m_i$. We see this condition is approximately
given by $m_T^2 >4$. 

Next we shall solve the valley equation.
Equation for $f$ is given by
\begin{equation}
f''+3 \cot \sigma \: f'-V_T''(\phi) f = -\lambda f.
\end{equation}
The regularity conditions are the same with that of $\phi$. Then, 
the general solution which satisfies the boundary conditions is given by
\begin{eqnarray}
f = \left\{ 
\begin{array}{ll}
\displaystyle{
\frac{1}{\sqrt{1-z^2}}\:G_F \:B^1_{\nu_{F \lambda}}(z)},&
\qquad -1 \leq z < z_{\lambda}, \\
\\
\displaystyle{
 \frac{1}{\sqrt{1-z^2}}\:G_T \:P^1_{\nu_{T \lambda}}(z)},&
\qquad z_{\lambda} \leq z  \leq 1,\end{array}
\right.
\end{eqnarray}
where
\begin{equation}
\nu_{T \lambda }=\sqrt{\frac{9}{4}+(m_T^2+\lambda)}-\frac{1}{2}, \:\:\:
\nu_{F \lambda }=\sqrt{\frac{9}{4}-(m_F^2-\lambda)}-\frac{1}{2}. 
\end{equation}
From the junction conditions, we obtain
\begin{equation}
G_F=\frac{P^1_{\nu_{T \lambda}}(z_{\lambda})}
{B^1_{\nu_{F \lambda}}(z_{\lambda})} G_T ,\:\:\:
G_F=\frac{P^2_{\nu_{T \lambda}}(z_{\lambda})}{B^2_{\nu_{F \lambda}}
(z_{\lambda})} G_T.
\end{equation}
The equation has solutions only if $z_\lambda$ satisfies the following
equation
\begin{equation}
P^1_{\nu_{T \lambda}}(z_{\lambda})B^2_{\nu_{F \lambda}}(z_{\lambda})
-P^2_{\nu_{T \lambda}}(z_{\lambda})B^1_{\nu_{F \lambda}}(z_{\lambda})=0,
\end{equation}
which determines the junction time $z_{\lambda}$. 
Next we solve the equation for $\phi$;
\begin{equation}
\phi''+3 \cot \sigma \: \phi'-V_T'(\phi)=-f.
\end{equation}
In the equation, $f$ acts as the source. We can see that the special 
solution is given by
\begin{equation}
\phi-\varphi_i=\frac{f}{\lambda}.
\end{equation}
Then the solution which satisfies the boundary condition is given by
\begin{eqnarray}
\phi_V = \left\{ 
\begin{array}{ll}
\displaystyle{
 \varphi_F+\frac{1}{\sqrt{1-z^2}} \:A_F \:B^1_{\nu_F}(z) 
+\frac{1}{\lambda \sqrt{1-z^2}} \:G_F \:B^1_{\nu_{F \lambda}}(z)},&
\qquad -1 \leq z < z_{\lambda}, \\
\\
\displaystyle{
\varphi_T+\frac{1}{\sqrt{1-z^2}} \:A_T \:P^1_{\nu_T}(z)
+\frac{1}{\lambda \sqrt{1-z^2}} \:G_T \:P^1_{\nu_{T \lambda}}(z)},&
\qquad z_{\lambda} \leq z  \leq 1.\end{array}
\right.
\end{eqnarray}
From the junction conditions, we obtain the coefficients 
\begin{eqnarray}
A_F &=& \sqrt{1-z_{\lambda}^2} 
\frac{P^2_{\nu_{T}}(z_{\lambda}) (\varphi_F-\varphi_T)}
{P^1_{\nu_{T}}(z_{\lambda})B^2_{\nu_{F}}(z_{\lambda})-P^2_{\nu_{T}}
(z_{\lambda})B^1_{\nu_{F}}(z_{\lambda})},
\nonumber\\
G_F &=& -\lambda \: \sqrt{1-z_{\lambda}^2}
\frac{\varphi_F P^1_{\nu_{T}}(z_{\lambda})B^2_{\nu_{F}}(z_{\lambda})
- \varphi_T P^2_{\nu_{T}}
(z_{\lambda})B^1_{\nu_{F}}(z_{\lambda})}{ B^1_{\nu_{F \lambda}}(z_{\lambda})
(P^1_{\nu_{T}}(z_{\lambda})B^2_{\nu_{F}}(z_{\lambda})-
 P^2_{\nu_{T}}(z_{\lambda})B^1_{\nu_{F}}(z_{\lambda}))}, \nonumber\\
A_T &=& \frac{B^2_{\nu_{F}}(z_{\lambda})}{P^2_{\nu_{T}}(z_{\lambda})} A_F, 
\:\:\:\:\:\:
G_T = \frac{B^2_{\nu_{F \lambda}}(z_{\lambda})}
{P^2_{\nu_{T \lambda}}(z_{\lambda})} G_F. 
\end{eqnarray}
Note that in this model the deformation of the configurations is
essentially determined by $z_{\lambda}$. If $z_{\lambda}=z_0$, $G_i$
becomes $0$, so $\phi_V=\phi_B$ as expected.

To ensure that the valley bounce plays a role instead of the true saddle
point solution, we must examine the fluctuations around the valley
bounce have one negative mode and give the imaginary part to the 
path-integral. The equation which determines the eigenmodes
is given by 
\begin{equation}
g_{ n}''+3 \cot  \sigma  \: g_{ n}'-V_T''(\phi)
g_{ n} = - \rho_{ n} g_{ n}.
\end{equation}
Then, the eigenvalue equation which determines the eigenvalue
$\rho_{n}$ of these eigenmodes becomes
\begin{equation}
P^1_{\nu_{T \rho_{n}}}(z_{\lambda})B^2_{\nu_{F \rho_{n}}}(z_{\lambda})
-P^2_{\nu_{T \rho_{n}}}(z_{\lambda})B^1_{\nu_{F \rho_{n}}}(z_{\lambda})=0.
\end{equation}
Note that, $\lambda$ is one of the solutions $\rho_{n}$.

We should treat separately the case in which the valley bounce exists 
around the top of the barrier and passes through only one parabola.
We put the solution for $f$ as 
\begin{equation}
f=\sum^{\infty}_{n=0}b_n \cos n \sigma, 
\end{equation}
then the equation for $f$ is rewritten as \cite{Jensen}
\begin{equation}
\sum^{\infty}_{n=0} \left( [(n-1)(n+2)-m_T^2-\lambda]b_{n-1}
-[(n+1)(n-2)-m_T^2-\lambda]b_{n+1} \right) \sin n \sigma =0.
\end{equation}
Thus, $b_n$ converges only when
\begin{equation}
\lambda=-m_T^2+n(n+3).
\end{equation}
We put the solution for $\phi$ as 
\begin{equation}
\phi-\varphi_T=\sum^{\infty}_{n=0}a_n \cos n \sigma, 
\end{equation}
then the solution for $\phi$ is given by
\begin{equation}
a_n=\frac{1}{\lambda}c_n.
\end{equation}
The eigenvalue of the eigenmode at the valley bounce is given by
\begin{equation}
\rho_{n}=-m_T^2+n(n+3).
\end{equation}

\subsection{The structure of the valley}
Using the analytic solution of $\phi$ and $f$, we show the structure of
the valley. Remember that we have rescaled the field as 
$\phi \to (M^2/m_T) \phi$ and $f \to (M^2/m_T) f$.
Following, for completeness, we consider the two types of the potential; 
(1)$m_T^2 >4$ and (2)$m_T^2<4$.  

(1)$m_T^2>4$ 

There exist two solutions in the Euclidean field equation; 
the CD solution and the HM solution. For example, we take $m_T^2=7$,
$m_F^2=2.2$ and $\eta=0.6 M^4$ (Fig.1). The behaviors of the CD solution 
$\phi(\sigma)$, the eigenmode with the negative eigenvalue $g_{-}(\sigma)$ 
and the scale factor $a(\sigma)$ are shown in Fig.2.
The CD solution has one negative eigenvalue 
$\rho_{CD,-}=-4.7$ and the smallest positive eigenvalue is 
$\rho_{CD,+}=3$. Since the CD solution gives the saddle-point 
of the path-integral, we analyze the valley trajectory which contains 
the CD solution. At the CD solution on the valley trajectory, $f=0$.
The valley bounce near the CD solution is obtained by deforming 
the CD solution; $\phi_V=\phi_{CD}+\triangle \phi_V$. 
The deformation $\triangle \phi_V$ 
is due to the source term  $f \neq 0 $ in the equation for $\phi$. 
The equation for $f$ is almost the same with that for the eigenmode 
$g$ at the CD solution. Thus, $\lambda$ is given by
$\lambda=\rho_{CD}+\triangle \lambda$. To ensure the action varies most
gently along the valley trajectory, $\rho_{CD}$ should be the eigenvalue
of the smallest value. In this case, $\rho_{CD}$ has one negative eigenvalue,
so we take $\lambda$ at the CD solution as $\lambda(\phi_{CD})=\rho_{CD,-}$
or $\lambda(\phi_{CD})=\rho_{CD,+}$. 

First examine the valley trajectory associated with the negative
eigenvalue ($\lambda(\phi_{CD})=\rho_{CD,-}$). The valley bounce obtained 
from the analytic results developed in the previous section is shown in 
the lower-panel of Fig.3. 
Since $\rho_{CD,-}$ is the lowest eigenvalue, 
 $f$ does not have a node. In the equation for $\phi$, $f$ acts as the
force. So, the valley bounce in this trajectory is 
obtained by deforming the CD solution adding a one-direction force $f$. If 
$f>0$, the valley bounce has a structure of the small bubble and 
if $f<0$ it has a structure of the large bubble \cite{Aoyama3}. 
We plot the action along this trajectory in lower-panel of Fig.4.  
The CD solution gives the maximum of the action. 

Next consider the valley trajectory associated with the 
smallest positive eigenvalue ($\lambda(\phi_{CD})=\rho_{CD,+}$). 
We show the valley bounce in this trajectory in the upper panel of
Fig.3. Since $\rho_{CD,+}$ is the next to the lowest eigenvalue, 
$f$ has one node. In the equation for $\phi$, $f$ acts as the mass term
of $\phi$. So, in this trajectory, the valley bounce is obtained by 
modifying the mass of the field $\phi$. It is known that the CD solution
is smoothly connected to the HM solution if one decreases the mass around 
the top of the barrier \cite{Samuel}. Then, it is expected that this
trajectory connects the CD solution and the HM solution. 
The action along this valley trajectory is shown in the upper-panel of 
Fig.4. Since the degeneracy occurs in $\lambda$, we take the horizontal 
coordinate as the 'norm' of the solution $\vert \Phi \vert 
=\sqrt{2 \pi^2 \int a^3 (\phi-\varphi_T)^2}$ \cite{Aoyama3}.
We see the CD solution is the minimum and the HM solution 
is the maximum of the action and these solutions are smoothly connected
on this trajectory as expected.

(2)$m_T^2<4$.  

The saddle point solution is the HM solution. For example we take $m_T^2=2$, 
$m_F^2=0.5$ and $\eta=0.1 M^4$ (Fig.5). The HM solution has one negative
eigenvalue $\rho_{HM,-}=-2$ and the smallest positive eigenvalue is given by
$\rho_{HM,+}=2$. The generic feature of the valley bounce is
understood by the simple analysis of the case in which the valley bounce
exists only in one parabola. First consider the valley trajectory associated 
with the negative eigenvalue. The solution of the valley equation is 
essentially has a form $f=\lambda (\phi-\varphi_T)=\mbox{const}$. 
This solution does not represent the tunneling, so we seek the trajectory
associated with the smallest positive eigenvalue 
$\lambda(\phi_{HM})=\rho_{HM,+}$. 
The solution of the valley equation is given by
$\phi-\phi_T \propto \cos \sigma$ and $f = \lambda (\phi-\phi_T)$ (Fig.6).
In this trajectory, the HM solution gives the minimum of the action 
(Fig.7). The action grows as the variation of the
field becomes large, but the increase is relatively gentle.

The fluctuations around the valley bounce should have 
one negative mode to ensure that the valley bounce plays a role instead
of the HM solution. The valley bounce has a lowest eigenvalue 
$\rho_{V,-} < \lambda(\phi_V)$, which is negative on this trajectory.
Since this is the unique negative eigenvalue, the gaussian integration of
the fluctuations around this valley bounce gives the imaginary part to
the path-integral. Then, the valley bounce contributes 
to the false vacuum decay and describes the creation of the bubble of
the true vacuum.

\section{An open universe from valley bounce}
\hspace{0.1cm}
Using the results developed so far, we will study a model for 
an open universe inside the bubble.
Since the radius of the bubble $R$ is small compared with the Hubble
horizon \cite{CL}, then the curvature scale is greater than the energy 
of the matter inside the bubble $\rho_M$ even if the whole energy
of the false vacuum is converted to the energy of the matter, 
$\rho_M/M_p^2 \sim H^2 <1/R^2$ \cite{Yokoyama1}.
Thus, we need the second inflation in the bubble. To realize the second
inflation inside the bubble, the field should roll slowly down the
potential. It requires that the curvature of the potential is small 
compared with the Hubble. To avoid the $ad$ $hoc$ fine-tuning of the 
potential, we will assume the requirement is satisfied 
for all region of the potential.
Since $m_T <H $, the solution of the Euclidean field equation
is given by the HM solution.
If the tunneling is described by the HM solution, the field appears at 
the top of the barrier. Then large fluctuations are generated 
because at the top of the barrier 
the field experiences the quantum diffusion rather than 
the classical potential force.
Fluctuations in this diffusion
dominated epoch make the inhomogeneous delay of the start of the
classical motion, thus make large fluctuations.

The above argument is based on the assumption that the HM solution 
describes the false vacuum decay. However, it seems to be 
possible that one of the
valley bounces describes the decay instead of the HM solution.
One of the grounds is that in the de Sitter spacetime,
the choice of the quantum states of the quantum fluctuations
above the false vacuum will affect the tunneling process considerably 
\cite{Rubakov}. In the de Sitter spacetime, one has to specify the
state of the quantum fluctuations of the field besides specifying the
classical vacuum which is the average value of the field i.e. $\varphi_F$.
The dominant configuration in the path-integral depends on
the initial state of the quantum fluctuations above the false vacuum. 
In case of the flat spacetime, if the initial state is of higher energy 
than the ground state, 
the dominant configuration is given by one of the
valley bounces instead of the bounce solution \cite{Aoyama3}. 
Thus it seems natural to consider the situation in which one of the
valley bounces describes the tunneling and inside  
the created bubble is our universe, although it is difficult to identify 
the quantum state corresponding to the valley bounce. 
Hence we take the presumption that one of the valley bounces 
describes the tunneling and inside the bubble created from the valley 
bounce is our universe.

A problem about the assumption that an individual valley bounce
describes the tunneling is the interpretation of the valley bounce
in the Lorentzian spacetime.
Our interpretation is that the valley bounce determines the initial 
condition of the tunneling field in the open universe.
Within the fixed background approximation, we can make analytic
continuation about the background geometry 
from the Euclidean de Sitter space (3) (see Fig.8).
By the analytic continuation
\begin{equation}
\tau = i(\rho-\pi/2),\:\: \sigma=\sigma,
\end{equation}
we obtain the Lorenzian de Sitter spacetime (Region II)
\begin{equation}
ds^2 = d \sigma^2+a(\sigma)^2 \left(- d\tau^2+\cosh^2 \tau
d \Omega^2 \right). 
\end{equation}
We take the nucleation surface at $\tau=0$.
Region II is almost covered by the
false vacuum. Then we assume the effect which modifies
the dominant configuration from the HM solution to the valley bounce
also modifies the classical motion of the field in this region.
The field obeys the equation analytically continued from the 
valley equation (14). The solution of the equation is given by 
the analytic continuation of the valley bounce.
On the other hand, because the bubble expands classically at a
velocity rapidly approaching the velocity of light, 
inside the expanding bubble is well described by the usual
classical equation of motion.
On the light-cone of the center of the bubble 
$\sigma=\pi$, the coordinate is singular $a(\sigma)=0$.
We continue to the interior of the light-cone (Region I) by
\begin{equation}
r=\tau+ i \frac{\pi}{2},\:\:\: t=i(\sigma-\pi).
\end{equation}
The resulting metric is given by
\begin{equation}
ds^2 = -d t^2+b(t)^2 \left(d r^2+\sinh^2 r d \Omega^2 \right), 
\end{equation}
The expanding bubble is homogeneous and isotropic on the hypersurface 
on the hyperbolic time slicing $t=const.$ 
The interior of the light-cone can be viewed as an open 
Friedman-Robertson-Walker universe with scale factor $b(t)$.
The initial condition of the
tunneling field on $t=0$ hypersurface is determined by 
the behavior of valley bounce at $\sigma=\pi$.
After that time, the evolution of the field 
is described by the classical field equation.

Under these assumptions, we will construct a model for an open inflation
in the simple model with the polynomial form of the potential.
We connect the linear potential at the point the field appears after the
tunneling $\phi=\phi_{\ast}$,
\begin{equation}
V(\phi)=V_{\ast}- \mu^3 (\phi-\phi_{\ast}), \qquad (\phi> \phi_{\ast}).
\end{equation}
We demand the potential and its derivative are connected smoothly
at the connection point $\phi_{\ast}$. Then we obtain
\begin{eqnarray}
V_{\ast} &=&\epsilon+\eta-\frac{1}{2}m_T^2(\phi_{\ast}-\varphi_T)^2, 
\nonumber\\
\mu^3 &=& m_T^2 (\phi_{\ast}-\varphi_T).
\end{eqnarray}
The initial conditions of the field in the open universe are 
given by the valley bounce
\begin{equation}
\phi(t=0)=\phi_0(z=1)=\phi_{\ast},\:\:\: \dot{\phi}(t=0)=0.
\end{equation}
The field evolves obeying the classical field equation;
\begin{equation}
\ddot{\phi}+3 \coth t \: \dot{\phi}+V'(\phi)=0,
\end{equation}
then the solution of $\phi$ satisfies 
\begin{equation}
\dot{\phi}(t)=\mu^3 \: \frac{\cosh^3 t-3 \cosh t +2}{3 \sinh^3 t}.
\end{equation}
In the small $t$ this behaves as $(1/4) \mu^3 t$.
The classical motion during one expansion time is given by
$\vert \dot{\phi} \vert H^{-1}$. 
On the other hand the amplitude of the
quantum fluctuations is given by $\delta \phi \sim H$. The curvature
perturbation ${\cal R}$ produced by the quantum fluctuations is approximately
given by the ratio of these two quantities; 
\begin{equation}
{\cal R} \sim \frac{\delta \phi}{\vert \dot{\phi} \vert H^{-1}}
\sim \frac{H^3}{\mu^3} \sim \frac{H^2}{m_T^2} \left(\frac{H}{\phi_\ast
-\varphi_T} \right).
\end{equation}
This should be of the order $10^{-5}$ from the observation of the 
cosmic microwave background (CMB) anisotropies.
If $\vert \phi_{\ast}-\varphi_T \vert < H$, as in the case the HM
solution describes the tunneling, ${\cal R} >1$ and the scenario cannot
work well. Fortunately, from Fig.7,
we see for appropriate $\lambda$, the valley bounce gives the initial 
condition as $\vert \phi_{\ast}-\varphi_T \vert \sim O(1)(M^2/m_T)$, 
which is larger than the Hubble if $M >H$. For this initial condition, 
the potential force works and the field rolls slowly down the potential 
according to eq.(50). We expect the curvature perturbation can be suppressed 
for the valley bounce

\section{Fluctuations in the open universe}
In this section, we will calculate the observable fluctuations in 
the open universe from the valley bounce based on a model 
described in the previous section.
We will assume the decay is described by one of the valley bounces 
$\phi_0$.  Then we take one specific $\lambda$ in the following 
calculations. 
Under this assumption, we calculate the observable fluctuations
in the open universe. 

\subsection{Fluctuations around the valley bounce}
We first calculate the fluctuations around the valley bounce in the
Euclidean spacetime.
The Euclidean action can be expanded around the valley bounce $\phi_0$ as 
\begin{equation}
S_E(\phi)=S_E(\phi_{0})+ \int d^4 x \sqrt{g} 
\frac{1}{\sqrt{g}}  \left.
\frac {\delta S_E}{\delta \phi} \right \vert_{\phi_{0}}
\delta \phi(\sigma)
+\frac{1}{2} \int \int d^4 x d^4 x' 
\left. \frac{\delta^2 S_E}{\delta \phi(x)
\delta \phi(x')} \right \vert_{\phi_{0}} 
\delta \phi(x) \delta \phi(x').
\end{equation}
Since the valley bounce does not obey the field equation, then the first
order derivative term does not vanish. This can be avoided by
constraining the space of the fluctuations to that orthogonal to the 
gradient of the action;
\begin{equation}
\int d^4 x \sqrt{g} \: \left. \left(\frac{1}{\sqrt{g}}
\frac{\delta S_E}{\delta \phi} \right \vert_{\phi_{0}} \delta \phi
\right)=0.
\end{equation}

We must calculate the physical observable like two-point correlation 
function in the bubble described by $\phi_0$. 
For example, consider the variance of the 
scalar field $\langle \phi^2 \rangle -\langle \phi \rangle^2$,
where $\langle \:\: \rangle$ is average over $\rho$ and $\Omega$.
Analytically continuing to the Lorentzian spacetime, 
this corresponds to the average over the space in open universe. 
Since the variables which depend only on $\sigma$ obey the relation 
\begin{equation}
\langle \phi_{0} \rangle=\phi_{0}, \:\:\:
\langle \delta \phi(\sigma) \rangle=\delta \phi(\sigma),
\end{equation}
we can show 
\begin{equation}
\langle \phi^2 \rangle -\langle \phi \rangle^2=
\langle \delta \phi(\sigma,\rho,\Omega)^2 \rangle 
-\langle \delta \phi(\sigma,\rho,\Omega) \rangle^2.
\end{equation}
So the observable in the open universe can be evaluated from 
inhomogeneous fluctuations which do not have $O(4)$ symmetric configurations.
In the de Sitter spacetime, inhomogeneous fluctuations are expanded 
by scalar harmonics 
\begin{equation}
\delta \phi (\sigma,\rho,\Omega)=\int dp \: S_p(\sigma)
 Y_{plm}(\rho, \Omega).
\end{equation}
The harmonics obeys the orthogonal relation between different $p^2$.
Now the gradient vector is given by $f(\sigma)$ which is the mode of 
$p^2=-1$. 
Then, the inhomogeneous fluctuations which depend on $\rho$ and $\Omega$
 ($p^2 \neq -1 $) are orthogonal to the gradient of 
the action $f(\sigma)$ automatically.

Fluctuations around the valley bounce $\phi_0$ obey the field equation
\begin{equation}
\int dx'^4 \left.
\frac{\delta^2 S_E}{\delta \phi(x)
\delta \phi(x')} \right \vert_{\phi_0} \delta \phi(x')=0.
\end{equation}
In Region II of the Lorenzian de Sitter spacetime;
\begin{equation}
ds^2 = d \sigma^2+a(\sigma)^2 \left(- d\tau^2+\cosh^2 \tau
d \Omega^2 \right),
\end{equation}
we assume background field obeys the equation of motion analytically
coninued from the valley equation. Then the fluctuations in 
Region II is obtained by the analytical continuation (45);
\begin{equation}
\tau = i(\rho-\pi/2),\:\: \sigma=\sigma.
\end{equation}
Thus we will solve the equation (58) analytically continuing to
Region II.
The procedure to solve the equation is the same with that was done 
in the previous works \cite{Cohn,Yamamoto,Turok1,Turok2}.
We will follow their calculations. 
Expanding the fluctuations as
\begin{equation}
\delta \phi (\sigma,\tau,\Omega)=\int dp \: S_p(\sigma)
 Y_{plm}(\tau, \Omega),
\end{equation}
we obtain the equation of the fluctuations
\begin{equation}
\left(\frac{\partial^2}{\partial \tau^2}+2 \tanh \tau 
\frac{\partial}{\partial \tau}+ \frac{l(l+1)}{\cosh^2 \tau} 
\right) Y_{plm}(\tau,\Omega)=-(1+p^2)Y_{plm}(\tau,\Omega),
\end{equation}
\begin{equation}
S_p''(\sigma)+3 \cot \sigma S_p'(\sigma)+
\left( \frac{1+p^2}{\sin^2 \sigma} -V_T''(\phi_0) \right)
S_p(\sigma)=0,
\end{equation}
where
\[
 V_T''(\phi_0) = \left\{ 
\begin{array}{ll}
m_F^2,&
\qquad 0 \leq \sigma < \sigma_{\lambda},  \\
\\
-m_T^2,&
\qquad \sigma_{\lambda} \leq \sigma \leq \pi, \end{array}
\right.
\] 
and $z_{\lambda}=-\cos \sigma_{\lambda}$. Here $\lambda$ is determined by 
$\phi_0$. 
Since the temporal coordinate $\tau$ is included in the harmonics $Y_{plm}$,
the choice of the solution $Y_{plm}$ specifies the vacuum.
This choice is related to the initial quantum state of the 
fluctuations. We will take this initial state as Bunch-Davis
vacuum. The equation for $S_p(\sigma)$ can be rewritten as 
\begin{equation}
\left(-\frac{d^2}{d u^2}+U(u) \right) \left(\frac{S_p(u)}{a(u)}\right)=p^2 
\left(\frac{S_p (u)}{a(u)}\right),
\end{equation}
where
\begin{equation}
a(u)=(\cosh u)^{-1},\:\:\: 
U(u) =\frac{ V''(\phi_0)-2}{\cosh^2 u}, \:\:\: \tanh u=-\cos \sigma=z.
\end{equation}
Since $U(u) \to 0$ as $u \to \pm \infty$, the modes are continuous for 
$p^2 >0$. For $u_{\lambda}< u$, 
the potential has a valley, then some discrete modes exist
for $p^2 <0$.

First consider the continuous modes. Positive frequency mode should
satisfy the Klein-Gordon normalization
\begin{equation}
 -i \int d z \:\:\: \left[
\cosh^2 \tau \:d \Omega
\left(
\delta \phi^+_{plm}  \:\: (\partial_{\tau} \delta \phi^{+ \ast}_{p'l'm'})
- \:\: (\partial_{\tau} \delta \phi^+_{plm} ) \:\: \delta 
\phi^{+ \ast}_{p'l'm'}
\right) 
\right]_{\tau=0}
=\delta(p-p') \delta_{ll'} \delta_{mm'}.
\end{equation}
For simplicity we consider s-wave. The normalized 
positive frequency mode function of the Bunch-Davis vacuum is given by 
\begin{equation}
\delta \phi_{ \pm p}^+(\sigma,\tau)
=S_{ \pm p}(\sigma) Q_p(\tau) ,\:\:
Q_p(\tau)=\frac{e^{\pi p /2} e^{-i p \tau}
-e^{-\pi p /2} e^{i p \tau}}{\sqrt{2 \sinh{\pi p}} 
\:\:\: \cosh \tau},
\end{equation}
where $S_p$ is normalized as
\begin{equation}
\int^{1}_{-1} dz \:\:\: 
S_p(z) S_{p'}^{\ast}(z)=\frac{1}{8 \pi \vert p \vert}
\delta (p-p').
\end{equation}
Using this mode function, the fluctuations can be expanded as
\begin{equation}
\delta \phi = \int^{\infty}_{0} dp \:\:\: \left[(
\delta \phi_p^+ \:\: \hat{\mbox{\boldmath$a$}}_p +
\delta \phi_{-p}^+ \:\: \hat{\mbox{\boldmath$a$}}_{-p} )
+(h.c) \right],
\end{equation}
where $\hat{\mbox{\boldmath$a$}}_p$ annihilates the Bunch-Davis vacuum. 
We take the initial fluctuations $S_p(z)$ at 
$z=-1$ as the Klein-Goldon normalized mode
$F_p^F(z)$ on $-1 \leq z \leq 1$, then we evolve this mode using the field
equation to $z=1$. The resulting mode function is 
\begin{eqnarray}
\tilde{S}_p(z)= 
\left\{ 
\begin{array}{ll}
F^F_p(z),& 
\qquad -1 \leq z < z_{\lambda}, \\
\\
\alpha_p F^T_p(z)+ \beta_p F^T_{-p}(z),&  
\qquad z_{\lambda} \leq z \leq 1,\end{array}
\right.
\end{eqnarray}
where 
\begin{equation}
F^i_p(z) =  \frac{1}{\sqrt{1-z^2}} \frac{1}{4 \pi \sqrt{\vert p \vert}}
\left(
a^i_+ \Gamma(1-ip)P^{ip}_{\nu_i}(z)-a^i_- \Gamma(1+ip)P^{-ip}_{\nu_i}(z)
\right),
\end{equation}
and
\begin{eqnarray}
a^i_{+}&=&\sqrt{\frac{1 + \sqrt{ 1-\vert C^i_2 \vert^2/C_1^{i2}}}{2}} 
,\:\:\:\:
a^i_{-}=\left(\frac{C_2^i}{\vert C_2^i \vert}  \right)
\sqrt{\frac{1 - \sqrt{1-\vert C^i_2 \vert^2/C_1^{i2}}}{2}}, \nonumber\\
C^i_1(p)&=& 2 \pi \left(1+\frac{\sin^2 \pi \nu_i}{\sinh^2 \pi p} 
\right)
,\:\:\:\:
C^i_2(p)= -2 \pi^2 \frac{\Gamma[1-ip]}{\Gamma[1+ip]}
\frac{\sin \pi \nu_i}{\sinh^2 \pi p}
\frac{1}{\Gamma[-ip-\nu_i] \Gamma[1-ip+\nu_i]}. \nonumber
\end{eqnarray}
Here, $\alpha_p$ and $\beta_p$ are determined by the junction conditions
at $z_{\lambda}$
\begin{eqnarray}
\alpha_p(z_{\lambda})
=\left. \frac{ F^F_p d_z F^T_{-p} - F^T_{-p}d_z F^F_p}
{F^T_p d_z F^T_{-p}- F^T_{-p} d_z F^T_{p}} \right 
\vert_{z_{\lambda}},
\nonumber\\
\beta_p(z_{\lambda})
=\left. \frac{-  F^F_p d_z F^T_{p} + F^T_{p}d_z F^F_p}
{F^T_p d_z F^T_{-p}- F^T_{-p} d_z F^T_{p}} \right 
\vert_{z_{\lambda}},
\end{eqnarray}
where $d_z=d/d z$.
$\tilde{S}_p(z)$ is not normalized on $-1 \leq z \leq 1$. 
The normalized mode function is given by 
\begin{eqnarray}
S_p(z)= 
\left\{ 
\begin{array}{ll}
b_+ F^F_p(z) - b_- F^F_{-p}(z),& 
\qquad -1 \leq z < z_{\lambda}, \\
\\
(b_+ \alpha_p - b_- \beta_{-p})F^T_p(z) 
+ (b_+ \beta_p - b_- \alpha_{-p} ) F^T_{-p}(z),&  
\qquad z_{\lambda} \leq z \leq 1,\end{array}
\right.
\end{eqnarray}
where
\begin{equation}
b_{+}=\sqrt{\frac{D_1}{D_1^2-\vert D_2 \vert^2}}
\sqrt{\frac{1 + \sqrt{ 1-\vert D_2 \vert^2/D_1}}{2}} 
,\:\:\:\:
b_{-}=
\left(\frac{D_2}{\vert D_2 \vert}  \right)
\sqrt{\frac{D_1}{D_1^2-\vert D_2 \vert^2}}
\sqrt{\frac{1 - \sqrt{1-\vert D_2 \vert^2/D_1}}{2}},
\end{equation}
and
\begin{eqnarray}
D_1(p)&=&\frac{1}{2}(\vert \tilde{\alpha}_p \vert^2 
+\vert \tilde{\beta}_p \vert^2+1), \:\:\:\:
D_2(p)= 
\tilde{\alpha}_p \tilde{\beta}_p + \frac{C^F_2}{2 C^F_1},
\nonumber\\
\tilde{\alpha}_p&=&\alpha_p a_+^T-\beta_p a_{-}^{T \ast}, \:\:\:\:
\tilde{\beta}_p= \beta_p a_+^T-\alpha_p a_{-}^{T}. 
\end{eqnarray}

Next consider the discrete modes. We put $p^2=-\Lambda^2$.  
The Bunch-Davis positive frequency mode is given by 
\begin{equation}
\delta \phi^+_{\Lambda lm}= S_{\Lambda}(\sigma) 
Y_{\Lambda lm}(\tau,\Omega),
\end{equation}
where
\begin{equation}
Y_{\Lambda lm}(\tau,\Omega)=\sqrt{\frac{\Gamma[\Lambda+l+1]
\Gamma[-\Lambda+l+1]}{2}} \frac{P^{-l-1/2}_{\Lambda-1/2}(i \sinh \tau)}
{\sqrt {i \cosh \tau}},
\end{equation}
and $S_{\Lambda}$ is normalized as
\begin{equation}
\int^{1}_{-1} dz \vert S_{\Lambda}(z) \vert^2 =1.
\end{equation}
From the regularity condition similar to the valley bounce, 
the solution is given by
\begin{eqnarray}
\tilde{S}_{\Lambda}(z)= \left\{ 
\begin{array}{ll}
\displaystyle{
\frac{\alpha_{\Lambda}}{\sqrt{1-z^2}} 
\left(P^{\Lambda}_{\nu_F}(z)+\beta_{\Lambda} P^{-\Lambda}_{\nu_F}(z)
\right)},&
\qquad -1 \leq z <z_{\lambda}, \\
\\
\displaystyle{
\frac{1}{\sqrt{1-z^2}} P^{-\Lambda}_{\nu_T}(z)},&
\qquad z_{\lambda}\leq z \leq 1,\end{array}
\right.
\end{eqnarray}
where
\begin{equation}
\beta_{\Lambda}=\frac{\sin \pi \nu_F}{\pi} \Gamma[1+\Lambda+\nu_F]
\Gamma[\Lambda-\nu_F].
\end{equation}
From the junction condition, $\alpha_{\Lambda}$ is given by
\begin{equation}
\alpha_{\Lambda}=
\left.
\frac{P^{-\Lambda}_{\nu_T}}{P^{\Lambda}_{\nu_F}
+\beta_{\Lambda} P^{-\Lambda}_{\nu_F}}  \right \vert_{z_{\lambda}},
\:\:\:\:
\alpha_{\Lambda}=
\left.
\frac{d_z P^{-\Lambda}_{\nu_T}}{d_z P^{\Lambda}_{\nu_F}
+\beta_{\Lambda} d_z P^{-\Lambda}_{\nu_F} }  \right \vert_{z_{\lambda}}.
\end{equation}
Then $\Lambda$ is determined by the equation
\begin{equation}
P^{-\Lambda}_{\nu_T}(d_z P^{\Lambda}_{\nu_F}
+\beta_{\Lambda} d_z  P^{-\Lambda}_{\nu_F})
=d_z P^{-\Lambda}_{\nu_T} (P^{\Lambda}_{\nu_F}
+\beta_{\Lambda} P^{-\Lambda}_{\nu_F}).
\end{equation}
The mode $\tilde{S}_{\Lambda}$ is not normalized. 
The normalized mode is given by
\begin{equation}
S_{\Lambda}(z)=N_{\Lambda} \: \tilde{S}_{\Lambda} (z),\:\:\:
N_{\Lambda}=\left(
\int^{1}_{-1} dz \vert \tilde{S}_{\Lambda}(z) \vert^2
\right)^{-1/2}.
\end{equation}

\subsection{Initial fluctuations in the open universe}

Fluctuations propagate into the interior of the light-cone 
$\sigma=\pi (z=1)$. Since the coordinate system (59) is singular 
on the light-cone, we make analytic continuation by
\begin{equation}
r=\tau+ i \frac{\pi}{2},\:\:\: t=i(\sigma-\pi).
\end{equation}
The resulting metric is given by
\begin{equation}
ds^2 = -d t^2+b(t)^2 \left(d r^2+\sinh^2 r d \Omega^2 \right), 
\end{equation}
where $b(t)=\sinh t$. Since the fluctuations exponentially expand 
during the second inflation in the bubble, 
the shortwavelength modes are relevant. 
The matching condition across the lightcone in the Minkowski 
limit is given by 
\begin{eqnarray}
F^T_p(\sigma)Q_p(\tau) &\to& \frac{-i}{2 \sqrt{2 p}}
\frac{1}{\sqrt{2 \sinh \pi p}} R_p(r)
(a^T_+ e^{\pi p/2} T_p (\eta)-a^T_- e^{-\pi p/2}T_{-p}(\eta)),
\nonumber\\
F^T_{-p}(\sigma) Q_{p}(\tau) &\to& \frac{-i}{ 2 \sqrt{2 p}}
\frac{1}{\sqrt{2 \sinh \pi p}} R_p(r)
(a^T_+ e^{-\pi p/2} T_{-p}(\eta)-a^{T \ast}_- e^{\pi p/2}T_{p}(\eta)),
\end{eqnarray}
where 
\begin{eqnarray}
T_p(\eta) &=& e^{-ip \eta- \eta}, \: e^{\eta}=\tanh(t/2), \nonumber\\
R_p(r) &=& \frac{1}{\sqrt{2} \pi} \frac{\sin p \: r}{\sinh r}.
 \end{eqnarray}
Note that $R_p(r)$ is the normalized scalar harmonics 
$R_p(r)=Y_{p00}(r)$, where
\begin{equation}
Y_{plm}(r,\Omega) = \frac{p \Gamma[ip+l+1]}{\Gamma[ip+1]} 
\frac{P^{-l-1/2}_{ip-1/2}(\cosh r)}
{\sqrt {\sinh r}} Y_{lm}(\Omega),
\end{equation}
and $Y_{lm}$ is the usual spherical harmonics.
Then, the extension to the general modes with $l \neq 0, m \neq 0$ is 
straightforwardly given by replacing $R_p(r)$ to $Y_{plm}(r,\Omega)$.
We obtain the fluctuations inside the bubble
\begin{eqnarray}
\delta \phi &=& -i \sum_{lm} \int^{\infty}_{0} dp
\frac{1}{2 \sqrt{2 p}} Y_{plm}(r,\Omega)
\frac{1}{\sqrt{2 \sinh \pi p}} \nonumber\\
&\times&
\left[ \left( 
(e^{\pi p/2} g_1(p,\lambda) T_p(\eta)
+e^{-\pi p/2} g_2(p,\lambda)T_{-p}(\eta) ) \hat{\mbox{\boldmath$a$}}_p 
\right.\right.
\nonumber\\
& + & \left.
(e^{-\pi p/2} g^{\ast}_1(p,\lambda) T_{-p}(\eta)
+e^{\pi p/2} g^{\ast}_2(p,\lambda) T_p(\eta))\hat{\mbox{\boldmath$a$}}_{-p} 
 \right)  \left.+ (h.c.) \right],
\end{eqnarray}
where
\begin{eqnarray}
g_1(p,\lambda)&=&a^T_+(b_+ \alpha_p(z_{\lambda})-b_-  \beta_{-p}(z_{\lambda}))
-a^{T \ast}_-(b_+ \beta_p(z_{\lambda})-b_-  \alpha_{-p}(z_{\lambda})),
\nonumber\\
g_2(p,\lambda)&=&a^T_+(b_+ \beta_p(z_{\lambda})-b_-  \alpha_{-p}(z_{\lambda}))
-a^{T}_-(b_+ \alpha_p(z_{\lambda})-b_-  \beta_{-p}(z_{\lambda})).
\end{eqnarray}
This is the initial condition of the fluctuations in the open universe.
The discrete mode can be treated in the same way.  We obtain the
positive frequency mode 
\begin{equation}
\delta \phi^+_{\Lambda}= N_{\Lambda}\: \frac{P^{-\Lambda}_{\nu_T}(\cosh t)}
{\sinh t} Y_{\Lambda lm}(r,\Omega).
\end{equation}
In the limit $t \to 0$, this becomes
\begin{equation}
\delta \phi^+_{\Lambda}=\frac{N_{\Lambda}}{2 \Gamma[1+\Lambda]}
T_{\Lambda}(\eta) Y_{\Lambda lm}(r,\Omega),
\end{equation}
where $T_{\Lambda}=e^{\Lambda \eta-\eta}$.
Using this mode function, the fluctuations can be expanded as
\begin{equation}
\delta \phi =\sum_i \sum_{lm} \delta \phi_{\Lambda_i}^+ 
\hat{\mbox{\boldmath$a$}}_{\Lambda_i} +(h.c).
\end{equation}
In Fig.9 we plot the solutions $\Lambda$
for $m_T^2 < 4$ (Case (2) in section 3.1). We also show the normalization
factor $N_{\Lambda}$.  
We find two solutions of $\Lambda$. We call the mode with 
$0<\Lambda_{sub}<1$ the subcritical mode and the one with
$1<\Lambda_{sup}$ the supercritical mode \cite{Tanaka}.
In the case the background 
solution is given by the HM solution, one supercritical mode with 
$\Lambda_{sup}=\nu_T=\sqrt{9/4+m_T^2}-1/2 >1$ exists. In the present
case, the mass changes $m_0^2$ to $-m_T^2$ at $z_{\lambda}$, another
subcritical mode appears. Note that in the case the CD solution describes
the tunneling, the supercritical mode corresponds to the wall
fluctuation mode $\delta \phi_{w} \propto \dot{\phi}_{CD}$ with 
$\Lambda_{sup}=2$. Although in the present case the correspondence cannot
be held, the behavior of the supercritical mode resembles the wall
fluctuation mode.

\subsection{Curvature perturbations in the open universe}
In this subsection we restore the Hubble scale $H$.
The field evolves with the classical field equation 
inside the bubble (Region I).  
We should match the fluctuations (89), (93) to the solution of the
field equation with the background field satisfying the classical
equation of the motion (50).
Fluctuations of the scalar field give rise to a metric 
perturbations in the open universe. So, we will consider the
evolution of the gauge invariant gravitational potential.
First consider the continuous modes \cite{Turok1}.
The evolution equation for the gauge invariant gravitational potential 
$\Phi$ is given by 
\begin{equation}
\Phi^{''}_p-\frac{6(1-e^{2 \eta})}{3-2^{\eta}} \Phi^{'}_p+
\left(p^2+5-\frac{4(3+e^{2 \eta})}{3-e^{2 \eta}}\right) \Phi_p=0.
\end{equation}
For small $t$, $\Phi_p$ behaves as
$T_{\pm p}(\eta) e^{2 \eta}$. For
general $t$, $\Phi_p$ behaves as
\begin{equation}
\Phi_p \sim 
T_{\pm p}(\eta) e^{2 \eta} 
\left(1-\frac{p \mp i}{3(p \pm i)} e^{2 \eta}\right).
\end{equation}
Furthermore for small $t$, this metric perturbation is related
to the fluctuations of the scalar field by
\begin{equation} 
\Phi_p \sim \frac{4 \pi G \mu^3}{(\mp i p +2)H^2} e^{2 \eta}
\:\: \delta \phi_p.
\end{equation}
The initial fluctuations of the scalar field are given in eq.(89).
Then, the metric perturbation generated 
during the second inflation is given by
\begin{eqnarray}
\Phi &=& -i \: \frac{4 \pi G \mu^3}{H} \sum_{lm} \int^{\infty}_{0} dp 
\frac{1}{2 \sqrt{2 p}} Y_{plm}(r,\Omega) 
\frac{1}{\sqrt{2 \sinh \pi p}} \nonumber\\
&\times&
\left[ \left( 
(e^{\pi p/2} g_1(p,\lambda) \tilde{T}_p(\eta)
+e^{-\pi p/2} g_2(p,\lambda) \tilde{T}_{-p}(\eta) ) 
\hat{\mbox{\boldmath$a$}}_p 
\right.\right.
\nonumber\\
& + &
\left.(e^{-\pi p/2} g^{\ast}_1(p,\lambda) \tilde{T}_{-p}(\eta)
+e^{\pi p/2} g^{\ast}_2(p,\lambda) \tilde{T}_p(\eta))
\hat{\mbox{\boldmath$a$}}_{-p} 
 \right)  \left.+ (h.c.) \right],
\end{eqnarray}
where 
\begin{equation}
\tilde{T}_{\pm p}(\eta)=T_{\pm p}(\eta)
\frac{e^{2 \eta}}{\mp i p +2} 
\left(1-\frac{p \mp i}{3(p \pm i)}e^{2 \eta} \right).
\end{equation}
The variable which has a normalization that
relates more directly to the density perturbation after reheating 
is given by ${\cal R}=16 \pi G (V^2/V^2_{,\phi})\Phi$.
We define the power spectrum of ${\cal R}$ by
\begin{equation}
\mbox{}_{BD} \langle 0 \vert {\cal R}(r,\eta) {\cal R}(r',\eta) \vert
0 \rangle_{BD}=\sum_{lm}
\int^{\infty}_{0} dp \:\:\: Y_{plm}(r) Y_{plm}(r')
P_{{\cal R}}(p,\eta),
\end{equation}
where $\hat{\mbox{\boldmath$a$}}_{p} \vert 0 \rangle_{BD}=0$. 
Taking the limit $\eta \to 0 (t \to \infty)$, we obtain
\begin{eqnarray}
P_{{\cal R}}(p,\lambda)&=&
P_{BD}(p) \times
\nonumber\\
&& \!\!\!\!\!\!\!\!\!\!\!\!\!\!\!\!\!\!\!\!\!\!\!\!\!
\left[\vert g_1(p,\lambda)\vert^2+ 
\vert g_2(p,\lambda) \vert^2 -\frac{1}{\cosh \pi p}
\left(\frac{p-i}{p+i} g_1(p,\lambda)
g_2^{\ast}(p,\lambda)+\frac{p+i}{p-i} g_1^{\ast}(p,\lambda) 
g_2(p,\lambda) \right) \right],
\end{eqnarray}
where
\begin{equation}
P_{BD}(p)=\left(\frac{3 H^3}{\mu^3}
\right)^2 \frac{\coth \pi p}{2 p(p^2+1)}.
\end{equation}
The power of the continuous modes in the logarithmic interval $p$ at 
$p \gg 1$ is given by
\begin{equation}
\lim_{p \to \infty} \frac{p^3}{2 \pi^2} 
P_{{\cal R}}(p,\lambda) = \frac{1}{4 \pi^2} \left(\frac{3 H^3}{\mu^3}
\right)^2 \sim 
\left(\frac{M_{\ast}^2}{M_p M} \right)^4 \left(\frac{H}{m_T} \right)^2.
\end{equation}
Here we use the fact the valley bounce gives the initial condition as
$\vert \phi_{\ast}-\varphi_T \vert \sim M^2/m_T$, then 
$\mu^3 =m_T M^2$. This quantity should be of the order $10^{-10}$ 
from the observation. This can be achieved by taking 
$(M^2_{\ast}/M) \ll M_p$. 
We show the dependence of $\lambda$ in $P_{{\cal R}}$ in Fig.10.

The discrete modes can be treated in the same way.
$\Phi_{\Lambda}$ generated from $\delta \phi_{\Lambda}$ is given by
\begin{equation}
\Phi_{\Lambda}= \sum_{lm} \frac{2 \pi G \mu^3 N_{\Lambda}}
{H \Gamma[1+\Lambda]}
\tilde{T}_{\Lambda}(\eta) Y_{\Lambda lm}(r_R,\Omega),
\end{equation}
where
\begin{equation}
\tilde{T}_{\Lambda}=T_{\Lambda}(\eta) \frac{e^{2 \eta}}{\Lambda+2}
\left(
1+\frac{1-\Lambda}{3(1+\Lambda)} e^{2\eta} \right).
\end{equation}
Taking the limit $\eta \to 0$, we obtain the power spectrum of ${\cal R}$
\begin{equation}
P_{{\cal R}}(\Lambda_i,\lambda)=
\left(
\frac{3 H^3}{\mu^3}
\right)^2 
\left(\frac{N_{\Lambda_i}(\lambda)}{\Gamma[2+\Lambda_i(\lambda)]} \right)^2.
\end{equation}

In some open inflation model, the contribution of the discrete modes gives 
the strong constraint on the model \cite{Sasaki}.
The supercurvature mode produces very large scale metric
perturbations and enhances the amplitude of the low multipoles of the 
CMB anisotropies. No evidence for such enhancement in the observed
spectrum implies that the contribution of these discrete modes must not 
dominate the contribution of the continuous modes.
Furthermore, if the amplitude of the supercurvature modes is large, the
universe is not open but quasi-open beyond the coherent length of the
supercurvatuer modes \cite{Garriga}. 
In our model, however, the last factor in the power spectrum 
$P_{{\cal R}}(\Lambda_i,\lambda)$ is $O(1)$ (see Fig.10), 
then there is no inconsistency with the observed CMB anisotropies and
the universe looks like an infinite open universe described 
by the valley bounce.

The harmlessness of the supercurvature mode can be deduced from
the analysis of the case the CD solution describes the tunneling and the 
thin-wall approximation can be used. In this case, the supercritical
mode is the wall fluctuation mode given by 
$\delta \phi_w =N_w (\dot{\phi}_{CD})$. Here the
normalization constant is given by $N_w=(\int d \sigma a(\sigma) 
\dot{\phi}_{CD}^2 )^{-1/2}$. Within the thin wall approximation, this
can be evaluated as $N_w \sim (R S_1)^{-1/2}$, where $S_1$ is the surface
tension of the wall and $R$ is the radius of the bubble.
The surface tension of the wall is estimated by $S_1 \sim m_T (\triangle
\phi)^2$, where $\triangle \phi$ is the scale of the variation of the
field during tunneling. The curvature perturbation generated from this 
wall fluctuation is given by ${\cal R} \sim (H/ \dot{\phi}_{CD})\delta 
\phi_w \sim H/ \sqrt{R S_1}$. Using $R \sim 1/H$ and $m_T \sim H$, 
this can be estimated as ${\cal R} \sim H/ \triangle \phi$. 
Thus, the thickness of the barrier the field passes during the tunneling 
determines the amplitude of the curvature perturbation generated from the
wall fluctuation mode. In the case the valley bounce
describes the tunneling, the supercritical mode can not be interpreted 
as the wall fluctuation mode. But the behavior of the supercritical mode
resembles that of the wall fluctuation mode. 
Thus we expect this analysis can be applied. Since the valley bounce gives 
$\triangle \phi \sim M^2/m_T$, the contribution of the supercritical
mode is suppressed. Then, the constraint from the discrete mode is not
strong in this model.

\section{Conclusion}
\hspace{0.1cm}

It is difficult to provide the model which solves the horizon problem
and at the same time leads to the open universe in the context of the
usual inflationary scenario. In the one bubble open inflationary
scenario, the horizon problem is solved by the first inflation and
the second inflation creates the universe with the appropriate $\Omega_0$.
Many works have been done within this framework of the 
scenario and it is recognized this scenario requires additional
fine-tuning \cite{Linde1,Linde2}. 
The defect comes from the fact that the requirement  
the curvature around the barrier should be larger than the Hubble scale 
to avoid large fluctuations contradicts to the requirement 
the curvature of the potential should be small to realize the second 
inflation inside the bubble. Additional constraint comes from the
fluctuations generated in the decay process, which can be observed 
and reject some models \cite{Sasaki}.  

In this paper we pointed out a possibility of constructing a model 
without these difficulties by reconsidering the tunneling 
process. If the curvature around the potential is small, 
the tunneling is described by one of a family of the almost saddle-point 
solutions
\cite{Rubakov}. 
This is because the true saddle-point solution, that is, the Hawking Moss
solution does not satisfy the boundary condition for the false vacuum decay.
A family of the almost saddle-point solutions generally forms 
a valley line in the functional space. We called the configurations on
the valley line valley bounces. 
To identify the valley bounces,  we formulated the valley method 
in the de Sitter spacetime and clarified the structure of the valley bounces. 
In this method the valley bounces can be 
identified using the fact the trajectory they form in the functional
space corresponds to the line on which the action varies
most gently.

Our assumption is that one of the valley bounces 
describes the creation of the bubble inside of which is our open universe
and determines the initial condition in the open universe.
Based on this assumption, we found that there occurs the second inflation
without the large fluctuations even if the curvature around the
barrier is small compared with the Hubble scale.

 The fluctuations of the tunneling field 
give rise to the metric perturbation. These can
be observed in our open universe. 
The fluctuations around the valley bounce which are orthogonal to the gradient
of the action are relevant to the observable. We calculated the power
spectrum of the metric perturbations generated in the second inflation
and found these fluctuations can be compatible with the observations.
In some models of the open inflation, the discrete mode of the
fluctuations gives strong constraint on the model. We showed this is not
the case in our model. Hence, using the valley bounce, we can solve the
problem which arises in the open inflationary scenario besides the usual 
fine-tuning of the inflationary scenario.  
Then the one bubble open inflation model can be constructed in the simple
model with the polynomial form of the potential.

Our conclusion is based on the assumption that an individual valley
bounce describes the tunneling and gives the initial condition of the
tunneling field in the open universe.
Although this assumption seems to be plausible, we note that 
further investigations are needed for the justification of 
this assumption.

\section*{Acknowledgements}
We have benefited from useful discussions with M.Sakagami and
A.Ishibashi. We are grateful to M.Sasaki for useful comments.
The work of J.S. was supported by Monbusho Grant-in-Aid No.10740118
and the work of K.K. was supported by JSPS Research Fellowships for
Young Scientist No.04687
\hspace{0.8cm}

\pagebreak

\newpage

\section*{Figure captions}

\begin{description}

\item[Fig.1] 
The piece-wise quadratic potential $V_T(\phi)$. Here, 
we take $m_T^2=7$, $m_F^2=2.2$ and $\eta=0.6 M^4$. \\

\item[Fig.2]
The behavior of the CD solution $\phi(\sigma)$, the eigenmode with the
negative eigenvalue $g_{-}(\sigma)$ and the scale factor $a(\sigma)$.
The potential is taken as in Fig.1.\\

\item[Fig.3]
The behavior of the valley bounce. The horizontal coordinate
is $\sigma$. The lower-panel shows the behavior of the valley bounce
on the trajectory associated with the negative eigenvalue and the upper-panel
shows the behavior of the valley bounce on the trajectory associated 
with the smallest positive eigenvalue. \\

\item[Fig.4]
The action along the valley trajectory. The lower-panel shows
the action of the trajectory associated with the negative eigenvalue. 
The horizontal coordinate is $\lambda$. The upper-panel shows
the action of the trajectory associated with the smallest positive 
eigenvalue. The horizontal coordinate is the norm of the filed 
$\Phi=\sqrt{\int d \sigma a(\sigma)^3 
\vert \phi(\sigma)-\varphi_T \vert^2}$.\\

\item[Fig.5]
The piece-wise quadratic potential $V_T(\phi)$. We take $m_T^2=2$, 
$m_F^2=0.5$ and $\eta=0.1 M^4$. \\

\item[Fig.6]
The action along the valley trajectory associated with smallest
positive mode. The potential is taken as in Fig.5.\\

\item[Fig.7]
The behavior of the valley bounce 
in the valley trajectory associated with lowest
positive mode. The upper-panel shows the behavior of $\phi$ and the
lower-panel shows the behavior of $f$. The potential is taken as in Fig.5.\\

\item[Fig.8]
Conformal diagram of the de Sitter spacetime. 

\item[Fig.9]
The solution of $\Lambda$. Two solutions are shown. One corresponds to the 
subcritical mode $0<\Lambda_{sub}<1$ and another corresponds to
the supercritical mode $1<\Lambda_{sup}$.
The corresponding valley bounces are shown in fig.7.

\item[Fig.10]
The power spectrum of the curvature perturbation around the
valley bounce (continuous mode). 
The corresponding valley bounces are shown in fig.7.

\end{description}

\newpage 

\begin{figure}
 \epsfysize=9cm  
 \begin{center}
  \epsfbox{fig1.eps }
 \end{center}
 \label{f1}
\caption{}
 \end{figure}

\begin{figure}
 \epsfysize=9cm  
 \begin{center}
  \epsfbox{fig2.eps }
 \end{center}
\caption{}
\label{f2}
 \end{figure}

\begin{figure}
 \epsfysize=20cm  
 \begin{center}
  \epsfbox{fig3.eps }
 \end{center}
\caption{}
 \end{figure}

\begin{figure}
 \epsfysize=20cm  
 \begin{center}
  \epsfbox{fig4.eps }
 \end{center}
\caption{}
 \end{figure}

\begin{figure}
 \epsfysize=9cm  
 \begin{center}
  \epsfbox{fig5.eps }
 \end{center}
 \caption{}
 \end{figure}

\begin{figure}
 \epsfysize=9cm  
 \begin{center}
  \epsfbox{fig6.eps }
 \end{center}
 \caption{}
 \end{figure}

\begin{figure}[h]
 \epsfysize=20cm  
 \begin{center}
  \epsfbox{fig7.eps }
 \end{center}
 \caption{}
 \end{figure}

\begin{figure}[h]
 \epsfysize=9cm  
 \begin{center}
  \epsfbox{fig8.eps }
 \end{center}
 \caption{}
 \end{figure}
\newpage

\begin{figure}[h]
 \epsfysize=9cm  
 \begin{center}
  \epsfbox{fig9.eps }
 \end{center}
 \caption{}
 \end{figure}
\newpage

\begin{figure}[h]
 \epsfysize=9cm  
 \begin{center}
  \epsfbox{fig10.eps }
 \end{center}
 \caption{}
\end{figure}

\newpage 
\end{document}